\documentclass[letter, oldversion]{aa} 
\usepackage{graphicx}
\usepackage{txfonts} 
\usepackage[]{natbib}
\bibpunct{(}{)}{;}{a}{,}{,}

\newcommand{\degree}{^{\circ}}

\begin{document}

   \title{GRB 070724B: the first Gamma Ray Burst localized by SuperAGILE and its Swift X-ray Afterglow}

   \author{E.~Del~Monte\inst{1}, M.~Feroci\inst{1}, L.~Pacciani\inst{1}, Y.~Evangelista\inst{1,2}, I.~Donnarumma\inst{1}, P.~Soffitta\inst{1},
           E.~Costa\inst{1}, I.~Lapshov\inst{1}, F.~Lazzarotto\inst{1}, M.~Rapisarda\inst{3},
           A.~Argan\inst{1}, G.~Barbiellini\inst{4,5}, M.~Basset\inst{4}, A.~Bulgarelli\inst{6}, P.~Caraveo\inst{7}, A.~Chen\inst{7},
           G.~Di~Cocco\inst{6}, L.~Foggetta\inst{4}, F.~Fuschino\inst{6}, M.~Galli\inst{8}, F.~Gianotti\inst{6}, A.~Giuliani\inst{7}, C.~Labanti\inst{6}, P.~Lipari\inst{2}, F.~Longo\inst{4,5},
           M.~Marisaldi\inst{6}, F.~Mauri\inst{9}, S.~Mereghetti\inst{7}, A.~Morselli\inst{10}, A.~Pellizzoni\inst{7}, F.~Perotti\inst{7},
           P.~Picozza\inst{10}, M.~Prest\inst{11}, G.~Pucella\inst{1}, M.~Tavani\inst{1,10}, M.~Trifoglio\inst{6}, A.~Trois\inst{1}, E.~Vallazza\inst{4}, S.~Vercellone\inst{7},
           V.~Vittorini\inst{1}, A.~Zambra\inst{12},
           P.~Romano\inst{13,14}, D.~N.~Burrows\inst{15}, G.~Chincarini\inst{13,14}, N.~Gehrels\inst{16}, V.~La~Parola\inst{17}, P.~T.~O'Brien\inst{18}, J.~P.~Osborne\inst{18},
           B.~Preger\inst{19,20}, C.~Pittori\inst{19,20}, L.~A.~Antonelli\inst{19,21}, F. Verrecchia\inst{19,20}, P.~Giommi\inst{19, 22}, L. Salotti\inst{22}}

   \offprints{E. Del Monte}

   \institute{INAF IASF Roma, Via Fosso del Cavaliere 100, I-00133 Roma, Italy\\ 
              \email{ettore.delmonte@iasf-roma.inaf.it}
              \and
              Dip. di Fisica, Universit\`a degli Studi di Roma ``La Sapienza'', P.le A. Moro 5, I-00185 Roma, Italy 
              \and
              ENEA UTS Fusione Tecnologie, Via E. Fermi 45, I-00044 Frascati (Rm), Italy 
              \and
              INFN Trieste, Padriciano 99, I-34012 Trieste, Italy 
              \and
              Dip. di Fisica, Universit\`a di Trieste, Via Valerio 2, I-34127 Trieste, Italy
              \and
              INAF IASF Bologna, Via Gobetti 101, I-40129 Bologna, Italy 
              \and
              INAF IASF Milano, Via E. Bassini 15, I-20133 Milano, Italy 
              \and
              ENEA C.R. ``E. Clementel'', Via don Fiammelli 2, I-40128 Bologna, Italy 
              \and
              INFN Pavia, Via Bassi, 6 I-27100 Pavia, Italy 
              \and
              Dip. di Fisica, Universit\`a degli Studi di Roma ``Tor Vergata'',  Via della Ricerca Scientifica 1, I-00133 Roma, Italy 
              \and
              Dip. di Fisica e Matematica, Universit\`a dell'Insubria, Via Valleggio 11, I-20100 Como, Italy 
              \and
              Consorzio Interuniversitario per la Fisica Spaziale, Viale Settimio Severo 63, I-10133 Torino, Italy 
              \and
              INAF, Osservatorio Astronomico di Brera, Via E.\ Bianchi 46, I-23807 Merate (LC), Italy 
              \and
              Universit\`a{} degli Studi di Milano-Bicocca, Piazza delle Scienze 3, I-20126 Milano, Italy 
              \and
              Department of Astronomy \& Astrophysics, Pennsylvania State University,      
              525 Davey Lab, University Park, PA 16802, USA
              \and
              NASA/Goddard Space Flight Center, Greenbelt, MD 20771, USA           
              \and
              INAF IASF Palermo, Via U.\ La Malfa 153, I-90146 Palermo, Italy 
              \and
              Department of Physics \& Astronomy, University of Leicester, LE1 7RH, UK     
              \and
              ASI Science Data Center, Via G.\ Galilei, I-00044 Frascati (Rm), Italy 
              \and
              INAF personnel resident at ASI Science Data Center 
              \and
              Osservatorio Astronomico di Roma, Via di Frascati 33, I-00040 Monte Porzio Catone (Rm), Italy 
              \and
              Agenzia Spaziale Italiana, Unit\`a Osservazione dell'Universo, Viale Liegi 26, 00198 Roma, Italy 
              }

   \date{\today}

\abstract{GRB 070724B is the first gamma ray burst localized by
SuperAGILE, the hard X-ray monitor aboard the AGILE satellite. The
coordinates of the event were published $\sim 19$ hours after the
trigger. The Swift X-Ray Telescope pointed at the SuperAGILE
location and detected the X-ray afterglow inside the SuperAGILE
error circle. The AGILE gamma-ray Tracker and Minicalorimeter did
not detect any significant gamma ray emission associated with GRB
070724B in the MeV and GeV range, neither prompt nor delayed.
Searches for the optical afterglow were performed by the Swift
UVOT and the Palomar automated 60-inch telescopes, resulting in no
significant detection. Similarly, the Very Large Array did not
detect any radio afterglow. This is the first GRB event associated
with an X-ray afterglow with a firm upper limit in the 100 MeV --
30 GeV energy range.}


\authorrunning {E. Del Monte et al.}
\titlerunning {GRB 070724B: the first Gamma Ray Burst localized by SuperAGILE and its...}
\maketitle
%

\section{Introduction}

The Italian AGILE satellite mission \citep{Tavani_et_al_2006} was
launched on 23 April 2007 from India to an equatorial orbit at
$\sim 550$ km altitude and $2.5 \degree$ inclination. The AGILE
payload includes two imaging instruments: the Gamma Ray Imaging
Detector (GRID), composed of a Silicon Tracker and a
Minicalorimeter (MCAL), sensitive in the 30 MeV -- 50 GeV energy
band, with $\sim 2.5$ sr field of view (FOV) and 15 arcmin source
location accuracy ($\geq 10\sigma$ detection), and SuperAGILE
\citep{Feroci_et_al_2007}, a one-dimensional coded aperture
instrument, based on silicon microstrip detectors, operating in
the nominal 18 -- 45 keV energy band, with a 2 $\times$ 1-D FOV of
$68 \degree \times 68 \degree$ and $\sim$1 -- 2 arcmin source
location accuracy for high S/N sources. Although the nominal upper
boundary of the SuperAGILE energy band is 45 keV, for sources with
a hard spectrum like the gamma ray bursts (GRBs) a significant
signal may be detected up to about 60 keV.

AGILE is expected to detect $\sim$10 -- 15 GRBs per year in the
SuperAGILE hard X-ray band and $\sim$5 -- 10 in the GRID gamma ray
band, only few of which will be localized by SuperAGILE due to its
smaller FOV. Approximately 1 GRB per week is currently detected by
MCAL in the 350 keV -- 2.8 MeV energy range. The three GRBs
detected by SuperAGILE during the first four months of nominal
operation (mid-July to mid-November is roughly consistent with the
expected rate. AGILE is equipped with an on-board GRB detection
and localization system, based on the SuperAGILE and MCAL
ratemeters (Del Monte et al. 2007; Fuschino et al. 2007). Upon
trigger, a SuperAGILE localization of the event is searched for
and the position is transmitted to Earth within a timescale of
minutes by using the ORBCOMM constellation of telecommunication
satellites \citep{Deckett_1993}. The commissioning phase of the
on-board GRB detection system was not complete at the time of GRB
070724B, which was then identified and localized during ground
standard quicklook operations.

In this Letter we report the properties of the first GRB localized
by SuperAGILE and of its X-ray afterglow discovered by the X-Ray
Telescope \citep[XRT; ][]{Burrows2005:XRT} aboard Swift.
Throughout this paper the quoted uncertainties are given at 90\%
confidence level for one interesting parameter (i.e., $\Delta
\chi^2 =2.71$) and times are referred to the SuperAGILE trigger
$T_0$ (i. e., $t=T-T_0$), unless otherwise specified.


\section{Observation of GRB 070724B}

\subsection{Prompt emission}

On 24 July 2007 at 23:25:08 UT (hereafter $T_0$) a gamma ray burst
(GRB 070724B) was detected by SuperAGILE. Due to the
unavailability of the SuperAGILE on-board GRB detection and
localization system and to a problem that delayed the telemetry
data transmission, the data processing could only start on ground
only $\sim13$ hours after the event. In addition, the mission was
still in its early phase, with the calibration of the SuperAGILE
astrometry mostly relying on the ground measurements and on a
sample of just three known X-ray sources detected in flight.

\begin{figure} 
\includegraphics[angle=90, width=9. cm]{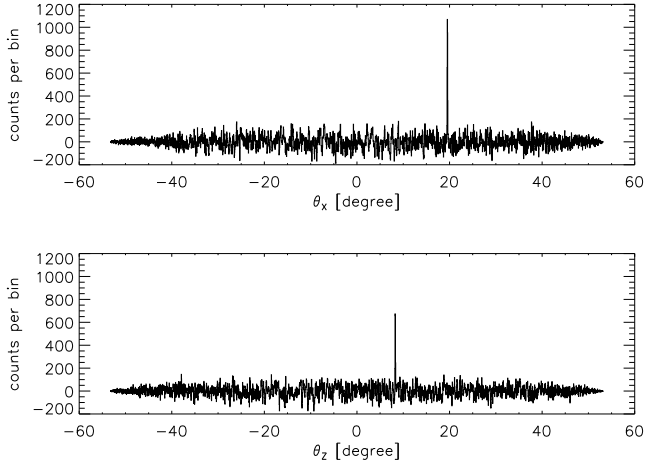}
\caption{SuperAGILE images of GRB 070724B. The offset angles
$\theta_X$ and $\theta_Z$ are refer to the satellite reference
frame, with a boresight pointing at $\mathrm{RA=01^h 17^m
30\farcs44^s}$, $\mathrm{Dec=+36 \degree 38' 45.7''}$ (equinox
2000).} \label{fig:SuperAGILE_images}
\end{figure}

\begin{figure}
\centering {\includegraphics[angle=0, width=8. cm]{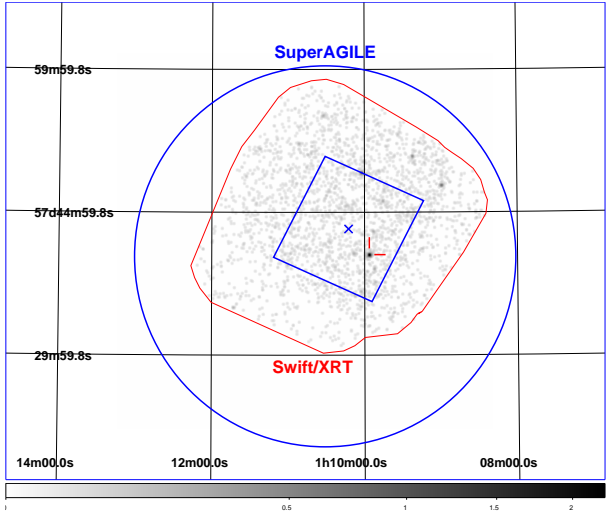}}
\caption{Swift/XRT image of the field of GRB~070724B, obtained
from the total $\sim16.5$\,ks PC mode data. The (red) polygon
highlights the shape of the sky covered by XRT, obtained with
several pointings. Also shown (red marks) is the XRT position, at
RA(J2000$)=01^{\rm h}$ $09^{\rm m}$ $56\fs33$,
Dec(J2000$)=+57^{\circ}$ $40^{\prime}$ $33\farcs0$ (90\% c.l.
error radius of $3\farcs9$). The (blue) circle is the early
20\arcmin{} SuperAGILE error circle, while the (blue) box inside
the XRT region is the SuperAGILE refined
$6\arcmin{}\times6\arcmin{}$ error box. The cross marks the
SuperAGILE refined position. } \label{grb070724B:fig:xrtmap}
\end{figure}

The first SuperAGILE GRB localization was derived by intersecting
the two one-dimensional images shown in Fig.
\ref{fig:SuperAGILE_images}, as $\mathrm{RA=01^h 10^m 31\fs0}$,
$\mathrm{Dec=+57 \degree 40' 23''}$ (equinox 2000) with an
uncertainty of 20 arcmin radius \citep{Feroci_GCN}, mostly
accounted for by the poor calibration of the absolute source
positioning available at the epoch of this detection. During the
science verification phase the SuperAGILE astrometry is calibrated
by means of a raster scan with the Crab Nebula and with the
detection of several other sources. With the status of calibration
available at the time of writing we derive the GRB refined
location as $\mathrm{RA=01^h 10^m 12\fs58}$, $\mathrm{Dec=+57
\degree 43' 14.9''}$ (equinox 2000), with an uncertainty of $\pm$6
arcmin (90 \% level), independently on each 1-D direction. The
preliminary and refined error boxes are shown in Fig.
\ref{grb070724B:fig:xrtmap}.

GRB 070724B also triggered the Konus-Wind \citep{Golenetskii_GCN},
Suzaku-WAM \citep{Endo_GCN} and INTEGRAL
SPI-ACS\footnote{http://isdc.unige.ch/cgi-bin/cgiwrap/$\sim$beck/ibas/spiacs/ibas\_acs\_web.cgi}
(V. Beckmann, private communication) experiments, all
participating in the InterPlanetary Network \citep[IPN, e.
g.][]{Hurley_et_al_1999}. Our GRB localization is consistent with
the IPN annulus (K. Hurley \& V. Pal'shin, private communication).

The detection of GRB 070724B is significant up to about 60 keV
energy. The GRB duration is $\sim 45$ s, with $T_{90}=(40 \pm 1)$
s in the 20 -- 60 keV energy band. The SuperAGILE light curve of
the prompt emission in the same energy band (Fig. \ref{fig:GRB
070724B_lc}) shows a multi-peaked structure with four main peaks
and internal variability. Considering the fraction of the
SuperAGILE detector illuminated by the event, the dead time
correction to the observed counts is smaller than 1 \%.

\begin{figure} 
    \includegraphics[angle=90, width=9. cm]{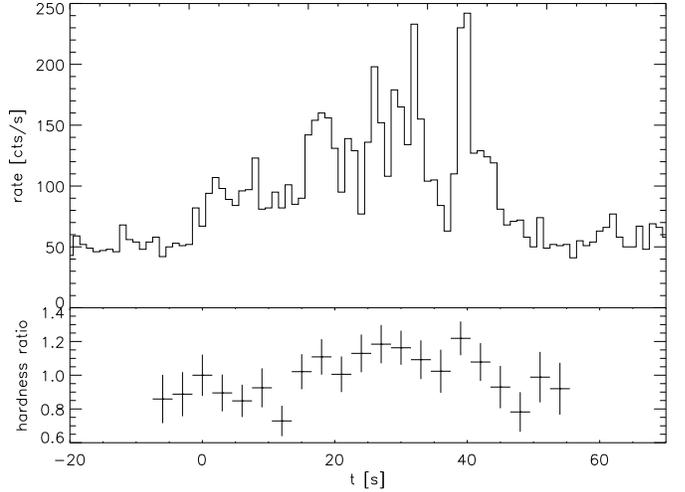}
    \caption{Upper panel: SuperAGILE lightcurve of GRB 070724B in the
    20 -- 60 keV energy band without background subtraction.
    The start time is 24 July 2007 at 23:25:08 UT.
    Lower panel: SuperAGILE hardness ratio of GRB 070724B.
    The lower energy band is 20 -- 26 keV, the higher 26 -- 60 keV,
    selected in order to have equivalent statistics.}
    \label{fig:GRB 070724B_lc}
    \end{figure}

The hardness ratio, estimated as the ratio of the counts in the 26
-- 60 keV and 20 -- 26 keV energy bands, selected in order to have
similar statistics, is shown in the lower panel of Fig.
\ref{fig:GRB 070724B_lc}. For comparison, the hardness ratio of
the Crab Nebula using the same energy bands is $\sim$0.6. The GRB
shows hints of spectral variability even in the narrow energy
range covered by SuperAGILE. The time-averaged spectrum in the
20~--~40 keV energy band, as observed by SuperAGILE, can be
described by a simple power law with photon index $\Gamma \sim
0.6$. Based on this analysis, the fluence in the 20 -- 40 keV
energy band is $\sim 5 \times 10^{-6}$ erg cm$^{-2}$ and the 1-s
peak flux (assuming the time-averaged spectral shape) is $\sim 4
\times 10^{-7}$ erg cm$^{-2}$ s$^{-1}$. Uncertainties on these
values cannot be taken lower than 50\%, due to the current very
early calibration status of the SuperAGILE experiment, for
off-axis events. The current status of the SuperAGILE response
matrix on-axis is such that the spectral parameters and flux of
the Crab Nebula are consistent with those from the literature
(e.g., Frontera et al. 2007) when a systematic uncertainty of
$\sim$15 \% is added to the model.

To search for gamma-ray emission from GRB 070724B we analyzed the
GRID data as follows. We first applied conservative cuts to the
GRID events in order to select only events that could be
recognized as celestial gamma-ray photons with a high confidence.
We only considered events of reconstructed energy larger than 100
MeV and arrival direction within $5 \degree$ from the coordinates
of the burst. This resulted in 0 events over a 50 s long time
window starting at $T_0$. The lack of events in this time interval
is statistically consistent with the expected background rate
($5.1 \times 10^{-4} \; \mathrm{cts \; s^{-1}}$) as determined
applying the same selection cuts to different time intervals. The
largest GRB fluence compatible (at the 99.5 \% confidence level)
with our observation of 0 photons is 0.02 $\mathrm{ph \; cm^{-2}}$
with $\mathrm{E>100}$ MeV (using the preliminary calibrations
currently available).

In the 350 keV -- 2.8 MeV energy band of the
AGILE-Minicalorimeter, no statistically significant counting rate
increase was found during the $\sim$50 s of the burst. The
corresponding 3-$\sigma$  upper limit is $\sim 4 \times 10^{-6} \;
\mathrm{erg \; cm^{-2}}$. This value has been calculated assuming
the  spectral shape as in Endo et al. (2007) extrapolated to 2.8
MeV, and the preliminary effective area obtained by MonteCarlo
simulations. The MCAL measurement is consistent with the lack of
detectable emission in the Konus-Wind 300 -- 1160 keV energy
band\footnote{http://www.ioffe.rssi.ru/LEA/GRBs/GRB070724\_T84307/}.
For this reason this GRB is reminiscent of the BATSE ``No High
Energy'' (NHE) type \citep{Pendleton_et_al_1997}. The GRB fluence
measured by Konus-Wind is ($1.80^{+0.04}_{-0.25}) \times 10^{-5}
\; \mathrm{erg \; cm^{-2}}$ (20 -- 500 keV energy band) with a
64-ms peak flux of ($2.17^{+0.34}_{-0.45}) \times 10^{-6} \;
\mathrm{erg \; cm^{-2} \; s^{-1}}$ in the same energy band and a
peak energy of $82 \pm 5$ keV \citep{Golenetskii_GCN}.


    \begin{figure}
        \resizebox{\hsize}{!}{\includegraphics[angle=270]{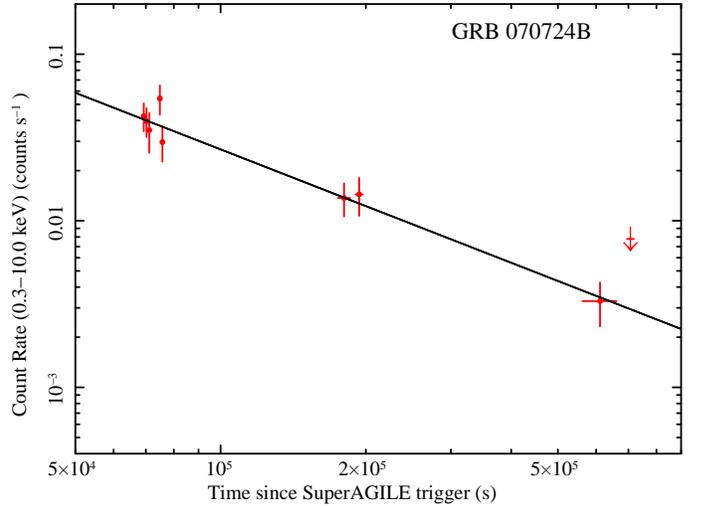}}
        \caption{Swift/XRT light curve of GRB~070724B, corrected for PSF losses, vignetting and
        background-subtracted. The solid line is the best-fit power-law model
        ($\alpha=1.13^{+0.23}_{-0.18}$).}
        \label{grb070724B:fig:xrtlcv}
    \end{figure}

\subsection{Afterglow emission}

The Swift 
data on GRB~070724B were collected as a Target of Opportunity (ToO)
observing campaign that commenced on 25 July 2007 at 18:32:39 UT,
19.1 hours after the burst trigger \citep{romano:gcn6669}. The
Swift/XRT data analysis was performed using standard procedures
\citep[see, e.g.][]{Romano et al 2006}.
%
The Swift/XRT observations were performed in photon-counting (PC)
mode, for a total on-source exposure of 16.5\,ks. Given the
$23\farcm6 \times 23\farcm6$ FOV, the observations were performed
so as to scan most of the area of the $20\arcmin$-radius
SuperAGILE initial error circle.
%
The only decaying source was found at
RA(J2000$)=01^{\rm h}$ $09^{\rm m}$ $56\fs34$,
Dec(J2000$)=+57^{\circ}$ $40^{\prime}$ $34\farcs3$, with an
uncertainty of $3\farcs9$ \citep[90\% c.l.\ radius;
][]{romano:gcn6703}.
Using the latest version of software and calibration, and
combining all available X-ray data, our final position is
(Fig.~\ref{grb070724B:fig:xrtmap}) RA(J2000$)=01^{\rm h}$ $09^{\rm
m}$ $56\fs33$, Dec(J2000$)=+57^{\circ}$ $40^{\prime}$ $33\farcs0$,
with an error radius of $3\farcs9$. 
This position is $4\farcm6$ from the early SuperAGILE position and
$3\farcm5$ from the refined one.

The decay and spectral indices are parameterized as $F(\nu,t)
\propto t^{-\alpha} \nu^{-\beta}$, where $F_{\nu}$ (erg cm$^{-2}$
s$^{-1}$ Hz$^{-1}$) is the monochromatic flux as a function of time
$t$ and frequency $\nu$; we also use $\Gamma = \beta +1$ as the
photon index, $N(E) \propto E^{-\Gamma}$ (ph keV$^{-1}$ cm$^{-2}$
s$^{-1}$).

Figure~\ref{grb070724B:fig:xrtlcv} shows the XRT light curve in
count rate in the 0.3--10 keV energy range, binned so that each
point has at least 30 counts, and a $>3$-$\sigma$ detection is
achieved. It exhibits a fading behaviour, with a power-law slope of
$\alpha=1.13^{+0.23}_{-0.18}$ ($\chi^2_{\rm red}=0.660/6$ d.o.f).

We fit the XRT spectrum of the first observation (4 ks) with XSPEC
(v11.3.2) using Cash statistics \citep{Cash1979}, which is more
appropriate than $\chi^2$ statistics given the low number of
counts (111), and spectrally unbinned data in the 0.3--10\,keV
energy range. We adopted an absorbed power law model with free
photon index $\Gamma$ and absorption using the photoelectric
absorption model {\tt tbabs} \citep{Wilms2000:tbabs} and
calculated the goodness of the fit via $10^4$ Montecarlo
simulations. We obtain $\Gamma=2.4\pm0.5$, a column density of
$(7\pm3)\times 10^{21}$ cm$^{-2}$, and a Cash statistic (C-stat)
of 371.6. We find that 76.40\% of the Monte Carlo realizations had
fit statistics lower than the observed C-stat. The column density
is
marginally in excess of the Galactic value \citep[$N_{\rm
H}^{\rm Gal}=3.1\times 10^{21}$ cm$^{-2}$; ][]{Kalberla2005:LAB}.
The conversion factor from count rate to unabsorbed flux in the
0.3--10\,keV energy range is 1 count s$^{-1}$ $\sim 1.3 \times
10^{-10}$~erg~cm$^{-2}$~s$^{-1}$, hence the unabsorbed
0.3--10\,keV flux for the spectrum at $T+24$\,hours is $\sim
4\times 10^{-12}$ ergs cm$^{-2}$ s$^{-1}$. This flux estimate
places GRB~070724B roughly at the average brightness for Swift
GRBs at this epoch of observation.
%

The Swift/UVOT \citep{Roming2005:UVOT} observations covered a
$17\arcmin \times 17\arcmin$ FOV, which is coaxial with that of
the XRT. The UVOT results were reported by \citep{chester:uvot}.
They refer to data collected starting from 2007 Jul 27 00:02:11
UT, 48.6\,hours after the burst trigger time, since the candidate
afterglow was outside the UVOT FOV during the first observation.
No uncatalogued sources were found in any of the UVOT observations
inside the refined Swift/XRT error circle for the candidate
afterglow, down to a 3-$\sigma$ limit in the co-added frames of
21.3\,mag in the White filter and 19.8\,mag in the $V$ filter.
These values were not corrected for Galactic extinction, which
corresponds to a reddening of $E_{B-V}=0.51$\,mag in the direction
of the burst \citep{Schlegelea98}.

\onltab{1}{        
 \begin{table*}
 \begin{center}
 \caption{Swift/XRT observation log.}
 \label{grb070724B:tab:alldata}
 \begin{tabular}{llllll}
 \hline
 \hline
 \noalign{\smallskip}
Sequence         & Start time  (UT)             & End time   (UT)                 & Net Exposure$^{\mathrm{a}}$  & Time since trigger$^{\mathrm{b}}$     \\
                 & (yyyy-mm-dd hh:mm:ss)         & (yyyy-mm-dd hh:mm:ss)        &(s)   & (s)    \\
 \noalign{\smallskip}
 \hline
 \noalign{\smallskip}
00020055001 &2007-07-25 18:32:39     &2007-07-25 20:34:52     & 3971    &68852   \\
00020055003 &2007-07-27 00:02:07     &2007-07-27 06:09:56     & 4002    &175019  \\
00020055004 &2007-07-31 11:27:41     &2007-07-31 17:57:58     & 3139    &561754  \\
00020055005 &2007-08-01 08:12:05     &2007-08-01 14:58:56     & 3302    &636418  \\
00020055006 &2007-08-02 00:31:49     &2007-08-02 07:02:56     & 2037    &695201  \\
 \noalign{\smallskip}
  \hline
  \end{tabular}
  \end{center}
  \begin{list}{}{}
  \item[$^{\mathrm{a}}$] The exposure time is spread over several snapshots (single continuous pointings at the target)
    during each observation.
   \item[$^{\mathrm{b}}$] Start time of the observation in seconds since the SuperAGILE trigger.
 \end{list}
\end{table*}
} 

\section{Discussion and conclusions}

GRB 070724B is the first gamma-ray burst localized by the
SuperAGILE experiment aboard the AGILE gamma-ray mission, still
performing its science verification phase. The field of view of
SuperAGILE is co-aligned with, and smaller than, that of the AGILE
GRID. Thus, although the SuperAGILE realtime GRB localization rate
is generally expected to be a minor addition to what is currently
provided (mostly) by Swift/BAT and INTEGRAL/IBIS, any GRB
localized by SuperAGILE will necessarily be in the GRID field of
view, and will have an associated measurement in the 30 MeV -- 50
GeV energy range. This makes any SuperAGILE GRB potentially
noticeable, because it may allow us, for the first time, to extend
up to the GeV energy band the study of a GRB with an associated
multi-wavelength afterglow and distance, measured through its
optical redshift. This would be a unique characterization of a
GRB, previously impossible. Indeed, for this purpose, AGILE was
equipped with on-board triggering and localization capabilities
and with a fast communication channel to distribute the few
arcmin-level SuperAGILE coordinates worldwide, with anticipated
delays of a few minutes. Unfortunately, the on-board procedure was
not active at the time of the first event, GRB 070724B, thus the
localization was delayed. The fast reaction by Swift allowed us to
discover the X-ray afterglow, but the UV, optical and radio
searches were carried out with large delays and did not set very
tight limits. Last but not least, the AGILE GRID did not detect
significant gamma-ray emission during the event. Thus, the
distance measurement of an AGILE/GRID burst is postponed until a
future opportunity, but GRB 070724B has nonetheless demonstrated
us the AGILE capabilities and allowed to observe, for the first
time ever, the X-ray afterglow of a GRB with a significant upper
limit on its gamma-ray emission.

The emission of photons in the $\sim$GeV energy range is required
by the commonly accepted GRB emission mechanisms in both external
and internal shocks, during the prompt phase \citep[for a review
see][]{Meszaros_2006}. Recently, the delayed GeV photons in GRB
(as in the case of GRB 940217 detected by EGRET) have been
theoretically correlated with the delayed X-ray flares detected by
Swift \citep[e.g.,][]{Burrows_et_al_2005b}, invoking, for example,
the inverse Compton scattering of X-ray flare photons by forward
shock electrons \citep{Wang_Li_Meszaros_2006}. The afterglow of
GRB 070724B was too weak (due to its late observation) to detect
delayed X-ray flares, and indication of time variability in the
early X-ray light curve, if any, is just marginal.

The phenomenology of GRBs in the $\sim$GeV energy range is still
poorly known, and not yet understood, due to the long gap between
the demise of the EGRET experiment and the launch of the
subsequent gamma-ray mission, AGILE. EGRET imaged high energy
photons associated with only five GRBs, also detected by BATSE
(Kwok et al. 1993; Hurley et al. 1994; Sommer et al. 1994; Schneid
et al. 1995). One other candidate was found by an independent
search \citep{Jones_et_al_1996}. Interestingly, the peak flux of
all the five BATSE GRBs was in the high flux tail of the BATSE log
N - log S distribution, approximately in the top 5 \%.

GRB 070724B did not show any detectable emission in the gamma ray
energy band. Indeed, combining this information with the
AGILE/Minicalorimeter and Konus observations
\citep{Golenetskii_GCN}, this event did not show any detectable
prompt emission above 300 keV, up to 30 GeV, and the two
properties (``NHE'' and lack of gamma emission) may be not
unrelated. In order to compare the properties of GRB 070724B with
those of the EGRET-detected events, we scaled the 64-ms Konus peak
flux to the BATSE standard 50 -- 300 keV energy band and obtained
$\sim9 \; \mathrm{ph \; cm^{-2} \; s^{-1}}$ ($\sim1.4 \times
10^{-6} \; \mathrm{erg \; cm^{-2} \; s^{-1}}$)\footnote[3]{Here we
assumed the spectral shape as provided by Konus for the
time-averaged spectrum. If we assume a power law shape with photon
index in the range 1--2, this number varies from $\sim 7$ to $\sim
12 \; \mathrm{ph \; cm^{-2} \; s^{-1}}$.}, thus positioning this
event in the brightest $\sim$10 \% of the BATSE peak flux
distribution. On the other hand, the 50-300 keV fluence of the
EGRET-detected GRBs was comparable to that of GRB 070724B ($\sim
10^{-5} \; \mathrm{erg \; cm^{-2} \; s^{-1}}$, except for the
noticeable case of 940217), but they all had large fluences, also
above 300 keV, where GRB 070724B was not detected. In particular,
the MCAL upper limit is $4 \times 10^{-6} \; \mathrm{erg \;
cm^{-2}}$ \citep[in agreement with the measurement in the 0.1 -- 1
MeV energy band by Suzaku, see][] {Endo_GCN}, a value
significantly smaller than the fluence measured by BATSE for the
GRBs detected by EGRET. Above 100 MeV, our upper limit of 0.02
$\mathrm{ph \; cm^{-2}}$ on the gamma-ray fluence compares to the
fluence ($>$100 MeV) of the brightest bursts seen with EGRET:
$\sim 0.039 \; \mathrm{ph \; cm^{-2}}$ for GRB 930131 (Sommer et
al. 1994) and $\sim 0.015 \; \mathrm{ph \; cm^{-2}}$ for GRB
940217 (Hurley et al 1994).

Thus, our observations and the properties of GRB 070724B are
consistent with the scenario suggested by the EGRET and BATSE
data, that the prompt gamma-ray emission correlates with the
energetics of the prompt emission at high energies (above 300 keV)
and not to those in the ``standard'' hard X-ray energy range
(e.g., 50 -- 300 keV), with only the brightest and hardest GRBs
having gamma-ray counterparts above the AGILE (and EGRET)
sensitivies. But our observations add a brand new piece of
information to this scenario: GRBs with undetectable gamma-ray
emission can however be associated with ``standard'' X-ray
afterglow emission, and possibly belong to the class of optically
dark events\footnote[4]{Adopting the method of the analysis in
\citep{De_Pasquale_et_al_2003}, the optical-to-X-ray flux ratio is
$f_{oX} \leq 2 \div 7$.}. In fact, an optical afterglow was not
detected by the Swift UVOT in the White and $V$ filters, nor by
the automated Palomar 60-inch telescope in the $I'$ filter,
yielding an upper limit of 21.0 mag \citep{Cenko_GCN} 31.3 hours
after the burst. Nor was the afterglow of GRB 070724B detected in
the radio band: a search by VLA at a frequency of 8.46 GHz on 5
August 2007 at 10.3 UT ($\sim 11.5$ days after the event) provided
a peak radio brightness of $-25 \pm 36 \, \mu Jy$
\citep{Chandra_GCN}.

Of course, the case of GRB 070724B does not allow us to establish
a new class of GRBs (yet ...), nor to draw any definite
conclusion, but still it allows us to make new correlations
between GRB properties, which were previously impossible. Even in
the unlikely situation that this will remain a single case, the
GRB emission models will need to face these new constraints from
now on, in the AGILE era.

\begin{acknowledgements}
      AGILE is a mission of the Italian Space Agency, with
co-participation of INAF (Istituto Nazionale di Astrofisica) and
INFN (Istituto Nazionale di Fisica Nucleare). This work was
partially supported by ASI grants I/R/045/04, I/089/06/0,
I/011/07/0 and by the Italian Ministry of University and Research
(PRIN 2005025417). INAF personnel at ASDC are under ASI contract
I/024/05/1. The authors warmly acknowledge the support by the team
of the InterPlanetary Network (IPN).
\end{acknowledgements}

\end{document}